\documentclass[12pt]{article}
\usepackage{amsmath,amssymb,amsfonts}
\usepackage{graphicx}
\usepackage{hyperref}
\hypersetup{colorlinks,linkcolor={blue},citecolor={blue},urlcolor={red}} 
\usepackage{cite}
\usepackage[english]{babel}
\usepackage{bm}
\usepackage[utf8]{inputenc}
\usepackage{listings}
\usepackage[T1]{fontenc}
\usepackage{lmodern}
\usepackage{caption}
\usepackage{floatrow}
\usepackage{subcaption}
\usepackage{mathrsfs}
\usepackage{enumerate}
\usepackage[normalem]{ulem}
\usepackage{color}
\usepackage{array}
\begin{document}

\title{\hfill\mbox{\small}\\[-1mm]
\hfill~\\[0mm]
       \textbf{Predicting Neutrino Mixing Angles Using Group Presentations}        }
\date{}
\author{\\[1mm] 
Carlos Alvarado$^{1\,}$\footnote{E-mail: {\tt calvara@dcpihep.com}}~,
Janelly Bautista$^{2\,}$\footnote{E-mail: {\tt jbautista11@ucol.mx }}~,~and Alexander J.~Stuart$^{1,2\,}$\footnote{E-mail: {\tt astuart@ucol.mx}}\\
\\[1mm]
  \textit{\small $^1$Dual CP Institute of High Energy Physics,}\\
  \textit{\small C.P.~28045, Colima, M\'exico}\\[3mm]
\textit{\small $^2 $Facultad de Ciencias-CUICBAS, Universidad de Colima,}\\
  \textit{\small C.P.~28045, Colima, M\'exico}\\[3mm]
   }
\maketitle

\vspace{0.5cm}

\begin{abstract}
\noindent
By assuming there exist three massive non-degenerate Majorana neutrinos, it is possible to describe neutrino mixing with a residual, unbroken discrete Klein subgroup of a larger spontaneously broken flavor symmetry group.   Motivated by forthcoming measurements of leptonic CP violation, we revisit this framework by applying  group presentation rules to it.   We develop a method that is able to reproduce all previous results in the literature and may also hint at a possible group theoretical origin of CP violation in the Klein symmetry elements. This is due to the explicit appearance of a phase in them. However, for the cases considered in this analysis, it turns out that this phase can be removed. Still,  this new method warrants further study. 
\end{abstract}
\thispagestyle{empty}
\vfill
\newpage
\setcounter{page}{1}

%%%%%%%%%%%%%%%%
\newpage
%%%%%%%%%%%%%%%%

%%%%%%%%%%%%%%%%%%%%%%%%%%%%%%%%%%%%%%%%%%%%%%%%%%%%%%%

\section{Introduction\label{sec:intro}}
The measurement of a sizable reactor mixing angle by the Daya Bay\cite{dayabay}, RENO\cite{reno}, and Double Chooz\cite{doublechooz} collaborations has opened the door for the measurement of leptonic CP violation.  Already the T2K \cite{T2KdeltaCP} and NO$\nu$A\cite{NOvadeltaCP} experiments are whittling away at the allowed values for the ``Dirac-type'' leptonic CP-violating phase $\delta$.  Additionally, global fits\cite{nufitpub,nufitweb, tortola, lisi1, lisi2} are  reducing the allowed parameter space for the Maki-Nakagawa-Sakata-Pontecorvo (MNSP) lepton mixing matrix, $U_{MNSP}=U_{e}^{\dagger}U_{\nu}$, where $U_e$ is the charged lepton mixing matrix and $U_{\nu}$ is the neutrino mixing matrix \cite{mnsp1, mnsp2, pdg}.  With all of this progress made on the experimental front of leptonic mixing, perhaps it may be a good time to revisit old paradigms which can predict leptonic mixing parameters when assuming three non-degenerate, nonzero Majorana neutrino masses.  By making this assumption, it is possible to implement  spontaneously broken discrete flavor symmetries which feature a residual Klein symmetry group in the neutrino sector at low energies\cite{kingluhnreview}.  The low-energy, residual Klein symmetry fixes the elements of $U_{MNSP}$ up to presumably Cabibbo-sized corrections\cite{cabibbohaze1, cabibbohaze2} (although it does not make any predictions for the Majorana phases \cite{majorana1,majorana2, majorana3}).

As a starting point for the discussion, we take the bottom-up approach for interpreting the Klein symmetry group elements in terms of low-energy leptonic mixing parameters as developed in Ref.~\cite{bottomup}.  Along with these elements, we incorporate   \textit{presentation rules/abstract definitions} for groups of particular interest, i.e., $A_4$, $A_5$, $S_4$, and $\Delta(96)$\cite{coxeter}.  This allows us to fix two angles and use the presentation rules to solve for the third.  The presentation rules can also be used to deduce new forms for the Klein elements.\footnote{For similar but not equivalent studies, see Refs.~\cite{Z2CP1i, Z2CP1ii, Z2CP1iii, Z2CP1iv, fonsecagrimus, hernsmirn1, hernsmirn2, hernsmirn3}.}

The remainder of this paper is structured as follows. In Section \ref{sec:method}, we review the Klein symmetry group framework and introduce the concept of presentations. In Section \ref{sec:genforms}, we define the explicit residual group symmetry elements for the neutrino and charged lepton sectors.  In Section \ref{sec:plato}, we look at a particular choice for the reactor and atmospheric mixing angles which greatly simplifies the analytic calculation of the solar angle.   In Section \ref{sec:applications}, we provide specific examples of our framework applied to the aforementioned groups of interest focusing on presentations with two generators, as well as applying the framework to a group that predicts a nonzero reactor mixing angle, i.e., $\Delta(96)$, and comment on its ability to predict a CP-violating phase.   In Section \ref{sec:conclusions}, we conclude.

%%%%%%%%%%%%%%%%%%%%%%%%%%%%%%%%%%%%%%%%%%%%%%%%%%%%%%%

%%%%%%%%%%%%%%%%%%%%%%%%%%%%%%%%%%%%%%%%%%%%%%%%%%%%%%%

\section{Background and Methodology\label{sec:method}}

%%%%%%%%%%%%%%%%%%%%%%%%%%%%%%%%%%%%%%%%%%%%%%%%%%%%%%%
It has become standard to define the residual neutrino flavor symmetry group $G_{\nu}$ as the Klein symmetry group $K_4$ such that
\begin{equation}\label{eq:symneut}
 G_{\nu}\cong K_4\cong Z^S_2\otimes Z^U_2, 
\end{equation}
where $S$ and $U$ are the generators for the two independent $Z_2$ symmetries.  Then it follows that the four elements of $K_4$ are $1$,  $S$, $U$, and $SU$.  In order to easily understand the breaking of the larger flavor symmetry group $G_f$ to this Klein subgroup, it is insightful to include $S$ and $U$ as two of the generators for the larger flavor symmetry  group $G_f$ as well.  To obtain the remaining generator of the flavor symmetry group $G_f$ we proceed by considering the residual symmetry in the charged lepton sector.  In the vast majority of models in the literature this group is 
\begin{equation}\label{eq:symcleps}
G_e\cong Z_n^T,
\end{equation}
where $T$ is the generator for this residual Abelian symmetry group in the charged lepton sector and $n\geq 3$.  With this assumption there now exist three possible generators for our flavor group $G_f$, i.e., $S$, $T$, and $U$.

%%%%%%%%%%%%%%%%%%%%%%%%%%%%%%%%

\subsection{Presentations with Two Generators\label{sec:2gen}}

%%%%%%%%%%%%%%%%%%%%%%%%%%%%%%%%

To begin we work with the simplest case and assume that the $U$ element can be expressed in terms of $S$ and $T$, e.g., in $A_5$, or that the $U$ is  a symmetry of the $S$ generator but not necessarily a group element, e.g., in $A_4$.  Then without loss of generality we can write our presentation as
	\begin{equation}\label{eq:STpres}
	\begin{aligned}
	S^2=T^n=(ST)^m=1.
	\end{aligned}
	\end{equation}
This above presentation is  a particular  form of a particular class of \textit{Dyck groups} defined by the presentation rules\cite{coxeter} 
\begin{equation}\label{eq:Dyck}
X^p=Y^q=(XY)^2=1.
\end{equation}
Notice that it is straightforward to show Eq.~\eqref{eq:STpres} implies Eq.~\eqref{eq:Dyck} by identifying $S=XY$, $T=Y^{-1}$, and $ST=X$, where $p=m$ and $n=q$.  This group is \textit{infinite} whenever\cite{coxeter} 
\begin{equation}\label{eq:infDyck}
(p-2)(q-2)=(m-2)(n-2)\geq 4.
\end{equation}
However, by assumption our flavor group $G_f$ is \textit{finite}.  This additional assumption puts restrictions on the values that $m$ and $n$ can take.  Furthermore, we will assume that neither $m$ nor $n$ (which are interchangeable by an isomorphism) is equal to 2 because if $m$ and/or $n$ is equal to 2 then we are left with a Dihedral group \cite{coxeter}.  Dihedral groups do not contain 3-dimensional irreducible representations.  Thus we will eliminate this case as well and take $n\geq 3$.\footnote{Note that this is also consistent with the original assumptions in Eq.~\eqref{eq:symcleps}.}  Then to obtain suitable \textit{finite} groups we arrive at the constraint in Eq.~\eqref{eq:infDyck},
\begin{equation}\label{eq:finDyck}
(p-2)(q-2)=(m-2)(n-2)\leq 3.
\end{equation}
where $m,n\geq 3$.  The solutions to this equality are straightforward to deduce.  They are
\begin{equation}\label{eq:solfinDyck}
(n,m)=(3,3),(4,3),(3,4),(5,3),(3,5).
\end{equation}
However by previous arguments, cf.~Eqs.~\eqref{eq:STpres}-\eqref{eq:Dyck}, we observe that an isomorphic group is generated when $n\leftrightarrow m$.\footnote{This isomorphism can easily be seen by letting $X=S$ and $Y=ST$ for both $S_4$ and $A_5$ and then relabeling  $Y$ as $T'$ to draw analogy with Eq.~\eqref{eq:STpres}. } Therefore, the solutions in Eq.~\eqref{eq:solfinDyck} only represent three groups, i.e., $A_4\cong(3,3)$, $S_4\cong(3,4)$, and $A_5\cong(3,5)$, where we have used the shorthand $(n,m)$ to denote a group given by the presentation in Eq.~\eqref{eq:STpres}.   However even though these $5$ solutions reduce down to only $3$ groups via isomorphism, we will show that the two different presentations for $S_4$ and $A_5$ generate \textit{different} predictions for the leptonic mixing parameters.  Nevertheless, before we analyze the phenomenology of the cases for the presentations given in Eqs.~\eqref{eq:STpres} and \eqref{eq:solfinDyck}, it is necessary to define  their explicit representations.

%%%%%%%%%%%%%%%%%%%%%%%%%%%%%%%%

\section{Explicit Generator Forms\label{sec:genforms}}

%%%%%%%%%%%%%%%%%%%%%%%%%%%%%%%%
Now that we have defined the cases in which we seek to explore in this work, i.e., those defined in Eqs.~\eqref{eq:STpres} and \eqref{eq:solfinDyck}, next we must provide the explicit forms for  $S$, $T$, and $U$.  We begin this discussion by considering the charged leptons.

%%%%%%%%%%%%%%%%%%%%%%%%%%%%%%%%%

\subsection{The Charged Lepton Sector and $\mathbf{Z_n}$}

%%%%%%%%%%%%%%%%%%%%%%%%%%%%%%%%%

As mentioned in Eq.~\eqref{eq:symcleps}, the standard charged lepton residual symmetry group is $Z_n$ ($n\geq 3$). Furthermore, (as is standard) we will work in a basis where the charged lepton mass matrix is diagonal using the convention in Ref.~\cite{bottomup} for $U_e$.  Thus, without loss of generality, it is possible to choose the 3-dimensional charged lepton $T$ generator as\footnote{We choose the following diagonal ordering to make connection to works on $A_4$($S_4$) where $T=$Diag$(1,\omega^2,\omega)$ with $\omega=e^\frac{2\pi i}{3}$, for example see Ref. \cite{kingluhnreview}.  However other orderings are possible depending on the presentation, e.g.,  Diag$(\omega^2,1, \omega)$, cf. Eq.~\eqref{eq:d96pres}.} 
\begin{equation}\label{eq:Tgen}
T_n=\left(
\begin{array}{ccc}
 1 & 0 & 0 \\
 0 & e^{-\frac{2\pi i k}{n}} & 0 \\
 0 & 0 & e^{\frac{2\pi i k }{n}}  \\
\end{array}
\right),
\end{equation}
where $k=1,\ldots, n-1$, and a subscript has been added to $T$ to denote its order.  Unlike the above representation for $G_e\cong Z_n^T$, the representation for the residual neutrino symmetry group $G_{\nu}\cong K_4\cong Z_2\otimes Z_2$ Klein symmetry group is more complicated as we will see in the next section.

%%%%%%%%%%%%%%%%%%%%%%%%%%%%%%%%%%%%%%%%%%%%

\subsection{The Neutrino Sector and $\mathbf{Z_2\otimes Z_2}$\label{subsec:neutsec}}

%%%%%%%%%%%%%%%%%%%%%%%%%%%%%%%%%%%%%%%%%%%%
By assuming that neutrinos are Majorana particles with nonzero and non-degenerate masses, it is straightforward to show that the residual flavor symmetry of the neutrino sector can be taken as the Klein symmetry group, which we will assume for the remainder of this paper.  The Klein symmetry group $ K_4\cong Z_2\otimes Z_2$ is an order-4 Abelian group.  As such, it has 4 elements: $\mathcal{I}, G_1, G_2,$ and  $G_3$.  In this work we choose the \textit{universal} representation for these elements as derived in Ref.~\cite{bottomup}, i.e.,
\begin{equation}
\begin{aligned}
G_1=&\left(
\begin{matrix}
 \left(G_1\right)_{11} & \left(G_1\right)_{12} & \left(G_1\right)_{13}  \\
 \left(G_1\right)_{12} ^* &\left(G_1\right)_{22} &\left(G_1\right)_{23}  \\
 \left(G_1\right)_{13}^*  &\left(G_1\right)_{23}^* & \left(G_1\right)_{33} \\
\end{matrix}
\right),\label{eq:newGs}
G_2=\left(
\begin{matrix}
\left(G_2\right)_{11} & \left(G_2\right)_{12} &\left(G_2\right)_{13} \\
  \left(G_2\right)_{12}^* &\left(G_{2}\right)_{22} &  \left(G_{2}\right)_{23} \\
 \left(G_2\right)_{13}^*&  \left(G_{2}\right)_{23}^*& \left(G_{2}\right)_{33}
\end{matrix}
\right),\\~~~~~~&~~~
G_3=\left(
\begin{matrix}
 -c'_{13} & e^{-i \delta } s_{23} s'_{13} & - e^{-i \delta }c_{23} s'_{13} \\
 e^{i \delta } s_{23} s'_{13} & s_{23}^2 c'_{13}-c_{23}^2 & - c_{13}^2  s'_{23} \\
 -e^{i \delta } c_{23} s'_{13} & - c_{13}^2  s'_{23} & c_{23}^2 c'_{13}-s_{23}^2 \\
\end{matrix}
\right),
\end{aligned}
\end{equation}
where $s_{ij}=\sin(\theta_{ij})$, $c_{ij}=\cos(\theta_{ij})$, $s'_{ij}=\sin(2\theta_{ij})$, $c'_{ij}=\cos(2\theta_{ij})$, and
\begin{equation}
\begin{aligned}
\left(G_1\right)_{11}&= c_{13}^2 c'_{12}-s_{13}^2,\\
\left(G_1\right)_{12}&= -2 c_{12} c_{13} \left(c_{23} s_{12}+e^{-i \delta } c_{12} s_{13} s_{23}\right),\\
\left(G_1\right)_{13}&= 2 c_{12} c_{13} \left(e^{-i \delta } c_{12} c_{23} s_{13}-s_{12} s_{23}\right),\\
\left(G_1\right)_{22}&=-c_{23}^2 c'_{12}+s_{23}^2 \left(s_{13}^2 c'_{12}-c_{13}^2\right)+\cos (\delta ) s_{13} s'_{12} s'_{23}, \\
\left(G_1\right)_{23}&= c_{23} s_{23} c_{13}^2+s_{13} \left(i \sin (\delta )-\cos (\delta ) c'_{23}\right) s'_{12}+\frac{1}{4} c'_{12} \left(c'_{13}-3\right) s'_{23},\\
\left(G_1\right)_{33}&=\left(s_{13}^2 c'_{12}-c_{13}^2\right) c_{23}^2-s_{23}^2 c'_{12}-\cos (\delta ) s_{13} s'_{12} s'_{23}, \\
\left(G_2\right)_{11}&= -c'_{12} c_{13}^2-s_{13}^2,\\
\left(G_2\right)_{12}&= 2 c_{13} s_{12} \left(c_{12} c_{23}-e^{-i \delta } s_{12} s_{13} s_{23}\right), \\
\left(G_2\right)_{13}&= 2 c_{13} s_{12} \left(e^{-i \delta } c_{23} s_{12} s_{13}+c_{12} s_{23}\right),\\
\left(G_2\right)_{22}&= c'_{12} c_{23}^2-s_{23}^2 \left(c_{13}^2+s_{13}^2 c'_{12}\right)-\cos (\delta ) s_{13} s'_{12} s'_{23}, \\
\left(G_2\right)_{23}&=e^{-i \delta } s_{13} s'_{12} c_{23}^2+\frac{1}{4} s'_{23} \left(2 c_{13}^2-c'_{12} \left(c'_{13}-3\right)\right) - e^{i \delta }  s'_{12} s_{13} s_{23}^2,
\\
\left(G_2\right)_{33}&=-c_{23}^2\left(c_{13}^2+s_{13}^2 c'_{12}\right) +s_{23}^2 c'_{12}+\cos (\delta ) s_{13} s'_{12} s'_{23} .
\end{aligned}
\end{equation}
In the literature, it is standard to identify $G_1=SU$, $G_2=S$, and $G_3=U$.  In this way these conventions can be easily mapped onto the existing literature.  We may now utilize these explicit forms of $S$, $SU$, $U$, and $T$ as tools  to probe the underlying leptonic flavor structure, similar to what was done in Ref.~\cite{bottomup}. Yet before that it would be useful to
highlight a few properties of the $G_i$ under some mixing angle choices which will greatly simplify the forthcoming calculations.

%%%%%%%%%%%%%%%%%%%%%%%%%%%%%%%%%%%%%%%%%%%%

\section{The Platonic Limit\label{sec:plato}}

%%%%%%%%%%%%%%%%%%%%%%%%%%%%%%%%%%%%%%%%%%%%
When building a model using explicit field content, generally the leading leptonic mixing predictions receive model-dependent corrections resulting from charged lepton corrections, renormalization group evolution, and canonical normalization considerations\cite{ CLRGECN1, CLRGECN2, CLRGECN3, CLRGECN4, genrge1, genrge2, genrge3, CN, CL1, CL2, CL3}.  These corrections can be used to shift the leading order predictions to more realistic values for the parameters.  In what follows we will work in the limit where the atmospheric angle is maximal and the reactor angle is zero.  This limit will be called the \textit{Platonic Limit} because it is in this limit that $G_1$, $G_2$, and $G_3$ take a particularly simple, idealized form, i.e.,
\begin{align}\label{eq:G1plato}
G^{plato}_1=\frac{1}{\sqrt{2}}\left(
\begin{array}{ccc}
 \sqrt{2} c'_{12} & -s'_{12} & -s'_{12} \\
 -s'_{12} & -\sqrt{2} c_{12}^2 & \sqrt{2} s_{12}^2 \\
 -s'_{12} & \sqrt{2} s_{12}^2 & -\sqrt{2} c_{12}^2 \\
\end{array}
\right),
\end{align}
\begin{align}\label{eq:G2plato}
G^{plato}_2=\frac{1}{\sqrt{2}}\left(
\begin{array}{ccc}
 -\sqrt{2} c'_{12} & s'_{12} & s'_{12} \\
 s'_{12} & -\sqrt{2} s_{12}^2 & \sqrt{2} c_{12}^2 \\
 s'_{12} & \sqrt{2} c_{12}^2 & -\sqrt{2} s_{12}^2 \\
\end{array}
\right),
\end{align}
\begin{align}\label{eq:G3plato}
G^{plato}_3=\left(
\begin{array}{ccc}
 -1 & 0 & 0 \\
 0 & 0 & -1 \\
 0 & -1 & 0 \\
\end{array}
\right).
\end{align}
Notice that it is in this limit $G_3$ takes its most simple and idealized (Platonic) form (as noted in Ref.~\cite{bottomup}).\footnote{ In Section \ref{sec:btmmixing}, this limit will be relaxed.}

Besides bringing $G_3$ to its most simple nontrivial form, the Platonic Limit also suggests a correspondence between $G_1$ and $G_2$ which will turn out to exist \textit{even when not working in this limit}.    Having said that, notice that it is possible to go from $G_2$ to $G_1$ by applying  the transformation 
\begin{equation}\label{eq:G2toG1}
\theta_{12}\rightarrow\theta_{12}\pm\pi/2
\end{equation}
because this transformation implies
\begin{equation}\label{eq:transG2G1} 
s_{12}\rightarrow \pm c_{12}, ~c_{12}\rightarrow \mp s_{12},~ c'_{12}\rightarrow ~ -c'_{12}.
\end{equation}   
Applying the shift  given in Eq.~\eqref{eq:G2toG1} to the $G_2$ in Eq.~\eqref{eq:G2plato} yields the $G_1$ given in Eq.~\eqref{eq:G1plato}.    Of course, applying the same transformation/operation again will transform back to $G_2$.  Therefore, taking the complement of the angle shifts from $G_2\leftrightarrow G_1$.   Thus we see that the transformation in Eq.~\eqref{eq:G2toG1}  (or equivalently transformations in Eq.~\eqref{eq:transG2G1})  forms another realization/representation of the action of $G_3$ on $G_1$ and $G_2$.

%%%%%%%%%%%%%%%%%%%%%%%%%%%%%%%%%%%%%%%%%

\section{Applications\label{sec:applications}}

%%%%%%%%%%%%%%%%%%%%%%%%%%%%%%%%%%%%%%%%%%%%
Now that we have the explicit forms for  $S$, $SU$, $U$, and $T$, cf. Eqs.~\eqref{eq:newGs} and \eqref{eq:Tgen}, respectively, we can now continue the discussion to view their power when it comes to presentations.  We begin by considering the three Dyck groups with two generators given in Eq.~\eqref{eq:solfinDyck}, i.e., $A_4$, $S_4$, and $A_5$.

%%%%%%%%%%%%%%%%%%%%%%%%

\subsection{$A_4$ and Tribimaximal Mixing\label{sec:A4}}

%%%%%%%%%%%%%%%%%%%%%%%%%%%

The group $A_4$ can be generated by two elements $S$ and $T_3$ subject to the following presentation rules, cf. Eq.~\eqref{eq:STpres} and  Eq.~\eqref{eq:solfinDyck}:
\begin{equation}\label{eq:A4pres}
S^2=T_3^3=(ST_3)^3=1.
\end{equation}
Recall that it is standard to let $S=G_2$, cf. Eq.~\eqref{eq:newGs}.  Additionally, $T_3=\text{Diag}(1,\omega^2,\omega)$, cf. Eq.~\eqref{eq:Tgen}.     
Thus with the identification of $S=G_2$ and $T_3=\text{Diag}(1,\omega^2,\omega)$, it is trivial to see that $S^2=T^3=1$. Yet, it is not clear if $(ST_3)^3$ can be 1. 

To analyze the condition under which  $(ST_3)^3=1$, first observe from $G_2$ in Eq.~\eqref{eq:newGs} that the product $ST_3$ will be daunting.  Thus, we will begin by taking the Platonic Limit, i.e., $\theta_{13}=0$ and $\theta_{23}=\pi/4$.  Then $ST_3$ becomes
\begin{align}\label{eq:STA4}
ST_3=
\left(
\begin{array}{ccc}
 -c'_{12} & \sqrt{2} c_{12} s_{12} \omega ^2 & \sqrt{2} c_{12} s_{12} \omega  \\
 \sqrt{2} c_{12} s_{12} & -s_{12}^2 \omega ^2 & c_{12}^2 \omega  \\
 \sqrt{2} c_{12} s_{12} & c_{12}^2 \omega ^2 & -s_{12}^2 \omega  \\
\end{array}
\right),
\end{align}
where $\omega=e^{\frac{2\pi i}{3}}$.   Solving for when $(ST_3)^3=1$ yields four distinct solutions for the solar mixing angle, i.e., 
\begin{align}\label{eq:A4sols}
\theta_{12}=\pm\cos^{-1}\left(\pm\sqrt{\frac{2}{3}}\right)\approx \pm 35.26^{\circ},~\pm 144.74^{\circ}.
\end{align}
Finally, the last key step is to notice that since the inputs are angles in the MNSP matrix, then they must live in the first quadrant by PDG convention\cite{pdg}.  Thus, the only solution to $(ST_3)^3=1$ when $\theta_{13}=0$ and $\theta_{23}=\pi/4$ is $\theta_{12}=\cos^{-1}\sqrt{2/3}\approx 35.26^{\circ}$.  Therefore, the presentation given for $A_4$ in Eq.~\eqref{eq:A4pres}  when working in the Platonic Limit yields a solar mixing angle $\theta_{12}=\cos^{-1}(\sqrt{2/3})$.  This solar mixing angle together with the Platonic Limit lead to tribimaximal (TBM) mixing\cite{tbm1, tbm2, tbm3, tbm4, tbm5, tbm6}, i.e.,
\begin{align}\label{eq:btmmixing} 
U_{\nu}^{TBM}=
\left(
\begin{array}{ccc}
 \sqrt{\frac{2}{3}} & \frac{1}{\sqrt{3}} & 0 \\
 -\frac{1}{\sqrt{6}} & \frac{1}{\sqrt{3}} & \frac{1}{\sqrt{2}} \\
 -\frac{1}{\sqrt{6}} & \frac{1}{\sqrt{3}} & -\frac{1}{\sqrt{2}} \\
\end{array}
\right).
\end{align}

%%%%%%%%%%%%%%%%%%%%%%%%%%%%%

\subsection{$A_5$ and Golden Ratio Mixing \label{sec:A5}}

%%%%%%%%%%%%%%%%%%%%%%%%%%%%%

The group $A_5$ can be generated by two elements $S$ and $T$ subject to the following presentation rules, cf., Eq.~\eqref{eq:STpres} and  Eq.~\eqref{eq:solfinDyck}:
\begin{equation}\label{eq:A5pres1}
S^2=T_3^3=(ST_3)^5=1
\end{equation}
or
\begin{equation}\label{eq:A5pres2}
S^2=T_5^5=(ST_5)^3=1.
\end{equation}
However, this presentation also requires $(ST_3)^5=1$.  Again,  we assume the Platonic Limit as a starting point to yield the same value as in Eq.~\eqref{eq:STA4}.  Yet this time we solve for the solar mixing under the condition $(ST_3)^5=1$ instead of the condition $(ST_3)^3=1$, as in the case of $A_4$.  This implies 8 (sometimes degenerate) solutions/conditions for the solar mixing angle, i.e.,
\begin{equation}\label{eq:A5sols1}
\theta_{12}=\pm\cos ^{-1}\left(\pm\sqrt{\frac{1}{6} \left(3\pm\sqrt{5}\right)}\right)\approx \pm 20.90^{\circ},~\pm 69.10^{\circ},~\pm110.90^{\circ},~\pm 159.10^{\circ}
\end{equation}
Unfortunately, this did not yield the known result\cite{A5GR1i, A5GR1ii, A5GR1iii} that $A_5$ yields the GR1 prediction of
\begin{equation}\label{eq:GR1pred}
\theta_{12}^{\text{GR1}}=\tan^{-1}\left(\frac{1}{\phi}\right)\approx 31.72^{\circ},
\end{equation}
or \begin{align}
U_{\nu}^{\text{GR1}}=\left ( \begin{array}{ccc} 
\sqrt{\frac{\phi}{\sqrt{5}}} & \sqrt{\frac{1}{\sqrt{5}\phi}} & 0\\ -\frac{1}{\sqrt{2}}\sqrt{\frac{1}{\sqrt{5}\phi}} & \frac{1}{\sqrt{2}}\sqrt{\frac{\phi}{\sqrt{5}} }& \frac{1}{\sqrt{2}}\\
-\frac{1}{\sqrt{2}}\sqrt{\frac{1}{\sqrt{5}\phi}} & \frac{1}{\sqrt{2}}\sqrt{\frac{\phi}{\sqrt{5}} }& -\frac{1}{\sqrt{2}}\\
\end{array} \right ).
\end{align}
where $\phi=\frac{1+\sqrt{5}}{2}$ is the Golden Ratio of Greek lore\cite{GR1i, GR1ii, A5GR1i, A5GR1ii, A5GR1iii} .

An attentive reader will realize that the results of Eq.~\eqref{eq:A5sols1} were not obtained when using $A_5$ in Ref.~\cite{A5GR1i}.  This can immediately be understood by observing that the authors of Ref.~\cite{A5GR1i} used a different presentation for $A_5$, i.e., the presentation given in Eq.~\eqref{eq:A5pres2}. Notice that the presentations given in Eq.~\eqref{eq:A5pres1} and Eq.~\eqref{eq:A5pres2} are isomorphic because they are related by algebraic relations in a similar fashion as Eqs.~\eqref{eq:STpres}-\eqref{eq:Dyck}.

Following the guidance of the form for the $T_5$ generator in Eq.~\eqref{eq:Tgen} we take
\begin{equation}\label{eq:A5Tgen}
T_5=\text{Diag}(1,\rho^4,\rho)
\end{equation}
where $\rho=e^{\frac{2\pi i}{5}}$.  Clearly this satisfies $T_5^5=1$ by construction.  Using this ``new'' $T=T_5$ generator, the product $ST_5$ can be made by again working in the Platonic Limit to simplify the calculation as well as work with good first approximations to the atmospheric and reactor mixing angles.  This yields
\begin{align}\label{eq:STA5}
ST_5=
\left(
\begin{array}{ccc}
 -c'_{12} & \rho ^4\sqrt{2} c_{12}  s_{12} & \rho\sqrt{2} c_{12}   s_{12} \\
 \sqrt{2} c_{12} s_{12} & -\rho ^4 s_{12}^2 & \rho c_{12}^2  \\
 \sqrt{2} c_{12} s_{12} & \rho ^4 c_{12}^2  & -\rho  s_{12}^2 \\
\end{array}
\right).
\end{align}
  Notice that (unsurprisingly) this is very similar to  Eq.~\eqref{eq:STA4}.  Then solving for the solar angle such that $(ST_5)^3=1$, cf.~Eq. ~\eqref{eq:A5pres2}, again yields 4 solutions: 
\begin{equation}\label{eq:A5sols2}
\theta_{12}=\pm\cos ^{-1}\left(\pm\sqrt{\frac{1}{10} \left(5-\sqrt{5}\right)}\right)\approx \pm58.28^{\circ}, \pm 121.72^{\circ}.
\end{equation}
Again observe that the angles in the above equation are different from the GR1 prediction given in Eq.~\eqref{eq:GR1pred}, i.e., $\theta_{12}\approx  31.72^{\circ}$.  However, they are complementary angles.  This results from the fact that even though the authors of Ref.~\cite{A5GR1i} let $S=G_2$, they chose a non-standard embedding for the right-handed charged leptons which led to  flipping of the second and third columns in the $G_1$ and  $G_2$ of Eqs.~\eqref{eq:G1plato}-\eqref{eq:G2plato}.  Then since overall minus signs on Klein symmetry elements do not matter, $G_1\leftrightarrow G_2$, 
 explaining the \textit{slight} difference between the results given in this work and that of Ref.~\cite{A5GR1i}.   Notice now, the presentation is not satisfied if $S=G_2$ and $T_5$ is as defined  in Eq.~\eqref{eq:A5Tgen}.  The $T_5$ needs to be changed to $\text{Diag}(1,\rho^2,\rho^3)$. Having concluded this discussion we now turn to the final Dyck group as defined in Eq.~\eqref{eq:solfinDyck}, i.e., $S_4$.

%%%%%%%%%%%%%%%%%%%%%%%%%%%%%

\subsection{$\mathbf{S_4}$ and Tribimaximal and Bimaximal Mixing\label{sec:S4}}

%%%%%%%%%%%%%%%%%%%%%%%%%%%%%

The group $S_4$ can be generated by two elements $S$ and $T$ subject to the following presentation rules, cf., Eq.~\eqref{eq:STpres} and  Eq.~\eqref{eq:solfinDyck}:
\begin{equation}\label{eq:S4pres3}
S^2=T_3^3=(ST_3)^4=1
\end{equation}
or 
\begin{equation}\label{eq:S4pres32}
S^2=T_4^4=(ST_4)^3=1
\end{equation}
Again for $S_4$ we begin with $T_3^3=1$.  Therefore, again $T_3=\text{Diag}(1,\omega^2,\omega)$ where $\omega=e^{\frac{2\pi i}{3}}$ and $T_4=\text{Diag}(1, -i,+i)$.  However, the $S_4$ requires $(ST_3)^4=1$ (instead of $(ST_3)^3=1$ as in the case of $A_4$).  Again, we work in the Platonic Limit of $\theta_{13}=0$ and $\theta_{23}=\pi/4$ to yield the exact same value as in Eq.~\eqref{eq:STA4}.  Yet this time the conditions for $(ST_3)^4=1$ must be solved instead of the conditions for $(ST_3)^3=1$, as in the case of $A_4$.
\begin{align}\label{eq:S4sols}
\theta_{12}=0,\pm\pi, \pm\cos^{-1}\left(\pm\sqrt{\frac{1}{3}}\right)\approx 0^{\circ},180^{\circ},~\pm 54.74^{\circ},~\pm125.26^{\circ}.
\end{align}
Then restricting to the first quadrant yields $\theta_{12}=0^{\circ}, 54.74^{\circ}$.  These angles are of course very disfavored and not tribimaximal mixing.  However, notice that  $54.74^{\circ}$ is the complementary angle of $35.26^{\circ}$, the solar mixing angle in tribimaximal mixing.  Thus, we should have taken $S=G_1$. 

Next, one may ask what the other presentation of $S_4$ yields.  Thus let us transform our presentation to Eq.~\eqref{eq:S4pres32}.
Then,
\begin{align}\label{eq:STS4}
ST_4=\left(
\begin{array}{ccc}
 -c'_{12} & -i \sqrt{2} c_{12} s_{12} & i \sqrt{2} c_{12} s_{12} \\
 \sqrt{2} c_{12} s_{12} & i s_{12}^2 & i c_{12}^2 \\
 \sqrt{2} c_{12} s_{12} & -i c_{12}^2 & -i s_{12}^2 \\
\end{array}
\right).
\end{align}
Solving $(ST_4)^3=1$ implies
\begin{align}\label{eq:S4sols2}
\theta_{12}=\pm\frac{\pi}{4},\pm\frac{3\pi}{4}.
\end{align}
Obviously $\theta_{12}=\frac{\pi}{4}$ is the only solution in the first quadrant.  However this yields a maximal solar and atmospheric angle with a vanishing reactor angle.  This is bimaximal mixing and $S_4$ is also known to generate this mixing pattern \cite{bimaximal1, bimaximal2, bimaximal3, bimaximal4, bimaximal5, S4bimax1, S4bimax2, S4bimax3}, i.e.,
\begin{align}\label{eq:bmmnsp}
U_{\nu}^{BM}=\frac{1}{2}\left(
\begin{array}{ccc}
 \sqrt{2} & \sqrt{2} & 0 \\
 -1 & 1 & \sqrt{2} \\
 -1 & 1 & -\sqrt{2} \\
\end{array}
\right).
\end{align}

Up until this point, all of the groups which we have considered are consistent with a vanishing reactor mixing angle.  However in the next section, we will show a group with a presentation that yields a nonzero reactor mixing angle.

%%%%%%%%%%%%%%%%%%%%%%%%%%%%%%%%%%%%%%%%%

\subsection{$\mathbf{\Delta(96)}$ and Bitrimaximal Mixing\label{sec:btmmixing}}

%%%%%%%%%%%%%%%%%%%%%%%%%%%%%%%%%%%%%%%%%
It is well known that $\Delta(96)\subset SU(3)$  is able to generate a nonzero reactor mixing angle\cite{BTmixing1, BTmixing2, delta96pres}.  Actually, there are many ways to present this $96$-element group.  Specifically, we focus on a presentation of $\Delta(96)$ which involves two generators, $U=G_3$ and $T$ \cite{delta96pres}, i.e.,
\begin{equation}\label{eq:d96pres}
U^2=T^3=(UT)^8=(U T^{-1}U T)^3=1,
\end{equation}
where now $T=\text{Diag}(\omega^2,1,\omega)$ in order to satisfy the above presentation for $U=G_3$ so that we can make connection with a nonzero reactor mixing angle.  Additionally, $G_3$ is the simplest Klein element and will help with the rather lengthy calculation due to the presentation, cf.~Eq.~\eqref{eq:d96pres}.  A quick numeric check shows that the following values for the mixing angles numerically satisfy the presentation rules given in Eq.~\eqref{eq:d96pres}, i.e., 
\begin{equation}\label{eq:BTMang}
\theta_{12}^{\text{BTM}}=\theta_{23}^{\text{BTM}}=\tan^{-1}(\sqrt{3}-1),\theta_{13}^{\text{BTM}}=\sin^{-1}((3-\sqrt{3})/6), \delta^{\text{BTM}}=0,
\end{equation}
leading to the bitrimaximal mixing matrix \cite{BTmixing1, BTmixing2, delta96pres}
\begin{align}\label{eq:btmnsp}
U^{BTM}_{\nu}=\frac{1}{\sqrt{3}}
\left(
\begin{array}{ccc}
 \frac{1}{2} \left(1+\sqrt{3}\right) & 1 & \frac{1}{2} \left(\sqrt{3}-1\right) \\
 -1 & 1 & 1 \\
 \frac{1}{2} \left(1-\sqrt{3}\right) & 1 & \frac{1}{2} \left(-1-\sqrt{3}\right) \\
\end{array}
\right),
\end{align}
proving that the novel method developed in this paper can also be applied to nonzero reactor mixing angles.  However, now it is possible to extend the reach of this method.  

In Eq.~\eqref{eq:btmnsp}, it was assumed that $\delta^{\text{BTM}}=0$.  However, this assumption can actually be relaxed, so that $\delta^{BTM}=\delta$, and $U=G_3^{\text{BTM}}$ becomes
\begin{align}\label{eq:UBTMCP}
G_3^{\text{BTM}}(\delta)= \frac{1}{\sqrt{3}}\left(
\begin{array}{ccc}
 -1-\frac{1}{\sqrt{3}} & \left(1-\frac{1}{\sqrt{3}}\right) e^{-i \delta } & -\frac{e^{-i \delta }}{\sqrt{3}} \\
 \left(1-\frac{1}{\sqrt{3}}\right) e^{i \delta } & -\frac{1}{\sqrt{3}} & -1-\frac{1}{\sqrt{3}} \\
 -\frac{e^{i \delta }}{\sqrt{3}} & -1-\frac{1}{\sqrt{3}} & 1-\frac{1}{\sqrt{3}} \\
\end{array}
\right).
\end{align}
Surprisingly, the above definition of $G_3^{\text{BTM}}(\delta)$ still satisfies the presentation rules given in Eq.~\eqref{eq:d96pres}, where now $U=G_3^{\text{BTM}}$ even with an arbitrary nonzero $\delta$.  Thus, we have possibly deduced a group, i.e., $\Delta(96)$, which furnishes a representation of the Klein elements that can predict a nonzero value for the CP-violating phase $\delta$.  Notice that the other generator of the residual Klein symmetry $S$ is as given in Footnote 8 of Ref.~\cite{delta96pres}, i.e., 
\begin{align}\label{eq:Sdelta96}
S^{BTM}(\delta)=U(UT)^4U(UT)^4=\frac{1}{3}\left(
\begin{array}{ccc}
 -1 & 2 e^{-i \delta } & 2 e^{-i \delta } \\
 2 e^{i \delta } & -1 & 2 \\
 2 e^{i \delta } & 2 & -1 \\
\end{array}
\right).
\end{align}
It is important to note that this $S$ is not equal to $G_2$ evaluated at the bitrimaximal mixing angles, but rather it is a complicated product of the $T=\text{Diag}(\omega^2,1,\omega)$  and $U=G_3^{BTM}$  generators which are evaluated at bitrimaximal mixing angles, cf.~Eq.~\eqref{eq:Sdelta96}.  However, it is interesting to note that
\begin{equation}\label{eq:correspondBTTB}
S^{BTM}(\delta=0)=\frac{1}{3}\left(
\begin{array}{ccc}
 -1 & 2 & 2 \\
 2 & -1 & 2 \\
 2 & 2 & -1 \\
\end{array}
\right)=G_2(\theta_{ij}^{TBM}),
\end{equation}
where $\theta_{ij}^{TBM}$ are the mixing angles associated with the tribimaximal mixing matrix.  The above element is actually equivalent to the well-known Klein element associated with the trimaximal middle columns of the tribimaximal leptonic mixing matrix prediction, cf.~Eq.~\eqref{eq:btmmixing}.\footnote{$S^{BTM}(\delta=0)=G_2(\theta_{ij}^{TBM})$ preserves the $(1,1,1)^T$ eigenvector.}

Finally with $G_3^{\text{BTM}}(\delta)$ and $S^{\text{BTM}}(\delta)$ it is possible to form the leptonic mixing matrix by looking at the eigenvectors associated with the +1 eigenvalues of these two Klein elements to reveal
\begin{align}\label{eq:btmdelta}
&U^{BTM}_{\nu}(\delta)=
&\left(
\begin{array}{ccc}
 \frac{1}{6} \left(3+\sqrt{3}\right) & \frac{1}{\sqrt{3}} & -\frac{1}{6} \left(\sqrt{3}-3\right) e^{-i \delta } \\
 \frac{3 \left(\sqrt{3}-1\right)+\left(2 \sqrt{3}-3\right) e^{i \delta }}{6 \sqrt{3}-15} & \frac{-3+\left(9-5 \sqrt{3}\right) e^{i \delta }}{6 \sqrt{3}-15}
   & \frac{1}{\sqrt{3}} \\
 \frac{1}{78} \left(12 \left(\sqrt{3}-4\right)+\left(9+\sqrt{3}\right) e^{i \delta }\right) & \frac{\left(3-2 \sqrt{3}\right) e^{i \delta }+3-3 \sqrt{3}}{6
   \sqrt{3}-15} & \frac{1}{6} \left(-3-\sqrt{3}\right) \\
\end{array}
\right)
\end{align}
which gives a Jarlskog Invariant\cite{jarlskog} equal to
\begin{equation}\label{eq:jarlskog}
J=\frac{1}{234} \left(9+\sqrt{3}\right) \sin (\delta ),
\end{equation}
showing that there may exist  CP violation for $\delta\neq 0,\pi$.  However, the interpretation that this phase generates physical CP violation is misleading.  Actually, it is erroneous to associate the phase $\delta$ with a physical CP-violating phase because this phase can be removed from $U=G_3$ by unphysical charged lepton rephasing\cite{bottomupext} but not from $G_1$ or $G_2$.   Nevertheless, there exists a basis for $\Delta(96)$ where the Clebsch-Gordan (CG) coefficients are real \cite{realCGdelta96i, realCGdelta96ii}, and if it depends on basis it cannot be physical\cite{holtlindCP}.  Finally, all representations of $Z_2\otimes Z_2$ are connected by unitary similarity transformations, so  it is possible to always remove the phases from the $G_i$ with said transformations\cite{bottomup}.  Recall that something similar happens in the case of $A_4$ when comparing CG coefficients in the Ma-Rajasekaran and Altarelli-Feruglio bases\cite{holtlindCP}.

Finally, observe that the above mixing matrix greatly simplifies when the CP-violating phase vanishes, i.e., 
\begin{align}\label{eq:btmdelta0}
U^{BTM}_{\nu}(\delta=0)=\frac{1}{\sqrt{3}}\left(
\begin{array}{ccc}
 \frac{1}{2} \left(1+\sqrt{3}\right) & 1 & \frac{1}{2} \left(\sqrt{3}-1\right) \\
 -1 & 1 & 1 \\
 \frac{1}{2} \left(1-\sqrt{3}\right) & 1 & \frac{1}{2} \left(-1-\sqrt{3}\right) \\
\end{array}
\right),
\end{align}
matching the leptonic mixing matrix for the original $U_{\nu}^{BTM}$ leptonic mixing matrix which assumed $\delta=0$, cf.~Eq.~\eqref{eq:btmnsp}.  Therefore, we have derived a representation for the $U$ generator of $\Delta(96)$ which is able to accommodate a nonzero reactor mixing angle.

%%%%%%%%%%%%%%%%%%%%%%%%%%%%%%%%%%%%%%%¸

\section{Conclusions \label{sec:conclusions}}

%%%%%%%%%%%%%%%%%%%%%%%%%%%%%%%%%%%%%%%%%
In this work, we have extended the applications  of the  bottom-up Klein elements derived in  Ref.~\cite{bottomup} by considering them alongside of a charged lepton generator and a set of presentation rules defining a group.  Making this extension defines a new framework where the appearance of the lepton mixing parameters is made explicit.  After fixing the reactor and atmospheric angles, the group presentation rules can be used to solve for the solar mixing angle.  Surprisingly,  this method verifies  existing results in the literature for the groups $A_4$, $A_5$, $S_4$, and $\Delta(96)$, and it shows that specific presentations can be characterized by specific sets of angles, cf.~Sections \ref{sec:A5} and \ref{sec:S4}.  Furthermore, by using this method it is possible to relate the $G_1$ and $G_2$ Klein elements by using the simple correspondence defined in Eq.~\eqref{eq:G2toG1}.  Finally, this method also allows for a nonzero reactor mixing angle as shown in the discussion of Section \ref{sec:btmmixing} for  bitrimaximal mixing and   $\Delta(96)$ (larger groups may be treated in a similar manner).  Although we have found that $\Delta(96)$ cannot accommodate a nonzero, physical CP-violating phase, this framework may be applied to discover other groups which can accommodate one.

\newpage

%%%%%%%%%%%%%%%%%%%%%%%%%%%%%%%%%%%%%%%%%%%%%%%%%%%%%%%

\section*{Acknowledgments}

%%%%%%%%%%%%%%%%%%%%%%%%%%%%%%%%%%%%%%%%%%%%%%%%%%%%%%%%%%%%%%%%%%%%%%% 

We thank A.~Aranda and M.~Pérez for helpful comments and suggestions.  We also thank E.~Díaz  and L.~Everett for their participation in early stages of this work, as well as for helpful comments and suggestions.   C.A. would like to thank the Facultad de Ciencias for their hospitality.  A.J.S. and J.B. would like to acknowledge partial support from CONACYT project CB2017-2018/A1-S-39470 (M\'exico).

\appendix

\end{document}